\documentclass[conference,review,authorversion]{IEEEtran} 

\AtBeginDocument{%
  }

\usepackage{listings}
\usepackage{tcolorbox}
\usepackage{caption}
\usepackage{subcaption}
\usepackage{tikz}
\usetikzlibrary{shapes.geometric, arrows}
\tikzstyle{startstop} = [rectangle, rounded corners, minimum width=3cm, minimum height=1cm,text centered, draw=black, fill=red!30]
\tikzstyle{io} = [trapezium, trapezium left angle=70, trapezium right angle=110, minimum width=3cm, minimum height=1cm, text centered, draw=black, fill=blue!30]
\tikzstyle{process} = [rectangle, minimum width=3cm, minimum height=1cm, text centered, draw=black, fill=white!30]
\tikzstyle{decision} = [diamond, minimum width=3cm, minimum height=1cm, text centered, draw=black, fill=green!30]
\tikzstyle{arrow} = [thick,->,>=stealth]

\newcommand*\circled[1]{\tikz[baseline=(char.base)]{
    \node[shape=circle, draw, inner sep=1pt, minimum size=8pt, font=\scriptsize] (char) {#1};}}
    
\newtcbox{\whiteshadowbox}[1][]{%
  enhanced, 
  center upper,
  fontupper=\large\bfseries,
  drop fuzzy shadow,
  boxrule=0.6pt,
   round corners,
  colframe=lightgray,
  colback=white!10,
  before={\begin{center}},
    after={\end{center}},
  #1%
}

\usepackage{balance}
\usepackage{verbatim}

\usepackage{fontawesome}
\usepackage{makecell}

\usepackage{stfloats}
\usepackage{textcomp}
\usepackage{booktabs, tabularx}
\usepackage{rotating}
\usepackage{soul}
\usepackage{pmboxdraw}
\usepackage{xcolor}
\def\SFii{\textcolor{black}{\textSFii\textSFx}}
\def\SFviii{\textcolor{black}{\textSFviii\textSFx}}

\def\SFxi{\textcolor{black}{\textSFxi}}

\usepackage{fancybox}
 \usepackage{multirow}

\usepackage{lipsum}
\usepackage{graphicx}
\usepackage{balance}
\usepackage{footnote}

\usepackage{array}

\usepackage{adjustbox}
\usepackage{threeparttable}

\usepackage{hyperref}
\hypersetup{
    colorlinks=true,
    linkcolor=black,
    anchorcolor=blue,
    filecolor=blue,      
    citecolor=black,
    urlcolor=black
}

\definecolor{brinkpink}{rgb}{0.98, 0.38, 0.5}

\newcommand{\phead}[1]{\vspace{1mm} \noindent {\bf #1}}

\definecolor{dkgreen}{rgb}{0,0.6,0}
\definecolor{gray}{rgb}{0.5,0.5,0.5}
\definecolor{mauve}{rgb}{0.58,0,0.82}
\definecolor{lightyellow}{RGB}{255,255,204}

\usepackage{array}
\newcolumntype{$}{>{\global\let\currentrowstyle\relax}}
\newcolumntype{^}{>{\currentrowstyle}}

\newcommand{\finding}[2]{
\begin{center}\vspace{-0.2cm}
    \resizebox{\linewidth}{!}{
\begin{tabular}{l!{\vrule width 0.5pt}p{0.9\columnwidth}}
    \makecell{{\LARGE \faLightbulbO}}  &\textbf{Finding {#1}.} 
{\em #2}
    \end{tabular}}
    
\end{center} 
}

\renewcommand{\textrightarrow}{$\rightarrow$}

\definecolor{hzcolor}{RGB}{10, 186, 181}

\begin{document}

\title{Discovery of Timeline and Crowd Reaction of Software Vulnerability Disclosures}

\author{
    \IEEEauthorblockN{Yi Wen Heng\textsuperscript{1}, Zeyang Ma\textsuperscript{1}, Haoxiang Zhang\textsuperscript{2}, Zhenhao Li\textsuperscript{3}, Tse-Hsun (Peter) Chen\textsuperscript{1}}
    \IEEEauthorblockA{\textit{\textsuperscript{1}Software PErformance, Analysis, and Reliability (SPEAR) lab, Concordia University, Montreal, Canada}}
    \IEEEauthorblockA{\textit{\textsuperscript{2}Queen's University}}
    \IEEEauthorblockA{\textit{\textsuperscript{3}York University}}

    \IEEEauthorblockA{he\_yiwen@encs.concordia.ca, m\_zeyang@encs.concordia.ca, haoxiang.zhang@acm.org}
    \IEEEauthorblockA{lzhenhao@yorku.ca, peterc@encs.concordia.ca}
}

\maketitle
\begin{abstract}

Reusing third-party libraries increases productivity and saves time and costs for developers. However, the downside is the presence of vulnerabilities in those libraries, which can lead to catastrophic outcomes. For instance, Apache Log4J was found to be vulnerable to remote code execution attacks. A total of more than 35,000 packages were forced to update their Log4J libraries with the latest version. Although several studies have been conducted to predict software vulnerabilities, the prediction does not cover the vulnerabilities found in third-party libraries. Even if the developers are aware of the forthcoming issue, replicating a function similar to the libraries would be time-consuming and labour-intensive. Nevertheless, it is practically reasonable for software developers to update their third-party libraries (and dependencies) whenever the software vendors have released a vulnerable-free version. In this work, our manual study focuses on the real-world practices (crowd reaction) adopted by software vendors and developer communities when a vulnerability is disclosed. We manually investigated 312 CVEs and identified that the primary trend of vulnerability handling is to provide a fix before publishing an announcement. Otherwise, developers wait an average of 10 days for a fix if it is unavailable upon the announcement. Additionally, the crowd reaction is oblivious to the vulnerability severity. In particular, we identified Oracle as the most vibrant community diligent in releasing fixes. Their software developers also actively participate in the associated vulnerability announcements.
     
\end{abstract}




\section{Introduction}
\label{sec:intro}

Modern software development often reuses third-party libraries to save time and costs incurred while developing from scratch. However, the presence of vulnerabilities (i.e., security flaws) could be catastrophic when the artifacts pose threats to their consumers. One way to assure software reliability is to refrain from vulnerabilities by keeping software library dependencies updated. 
However, a study showed that developers ceased to support and maintain dependency when a software development lifecycle ended~\cite{kula}, citing a lack of awareness of the new release. 
Another study~\cite{wallace} found that developers struggle with determining the most appropriate version when upgrading their dependencies.

The number of vulnerabilities has been steadily increasing since 2017~\cite{cve-distribution}. This trend is particularly concerning because vulnerabilities in widely used software libraries can create ripple effects of security risks across numerous dependent applications.
In November 2021, Apache Log4J was found to be vulnerable to remote code execution attacks, and more than 18,450 packages using the vulnerable version were affected~\cite{log4j-advisory-1}. 
Nevertheless, the first release of Apache's fix did not fully address the vulnerability. As a result, 18,560 packages needed to be updated twice to finally use a fixed version of the dependency~\cite{log4j-advisory-2}. 
A total of more than 35,000 impacted packages are roughly equivalent to 8\% of the total packages in the Maven ecosystem. Therefore, the magnitude of this impact is considered large as the average ecosystem impact of advisories affecting Maven Central is 2\%, with a median of less than 0.1\%~\cite{google-log4j}.

Existing literature has contributed to predicting software vulnerabilities using text analysis~\cite{software-vulnerability-prediction, software-vulnerability-prediction-2}, trend analysis with topic models~\cite{TrendAnalysis}, and machine learning~\cite{Yosifova2021PredictingVT}. However, vulnerability databases often lack sufficient data to create robust prediction models~\cite{nvd-not-good-enough-example, nvd-not-good-enough-example2} that accurately reflect how vulnerabilities evolve throughout their real-world lifecycle. Li et al.~\cite{Li2023TheAO} compiled a list of popular vulnerability databases that release official security policies and standards. However, these databases do not fully capture the real-world dynamics of vulnerabilities. Thus, we aim to address this gap by empirically investigating real-world vulnerability data to better understand the dynamics of how vulnerabilities evolve over time and how they are managed in practice.

This study aims to uncover the lifecycle of vulnerability management. We manually examine the timeline and responses from software vendors and the developer community following the disclosure of software vulnerabilities, particularly in third-party libraries commonly used in software development. Vulnerabilities in these libraries can lead to security risks across dependent applications, making it essential to understand how these issues are identified, fixed, and disclosed. 
More specifically, our analysis focused on real-world responses to disclosed vulnerabilities and the time taken by software vendors to release fixes. We built a dataset containing 735 Java-related CVEs reported after 2017, and conducted a detailed study on a random sample of 312 CVEs. Data collection involved manual examination of CVE records, GitHub repositories, and reference links to construct a timeline of key events in each vulnerability’s lifecycle.
 
Our study found that most vendors release fixes before publicly announcing vulnerabilities. We also observed that software vendors often upload their fixes across multiple platforms (but with time gaps), with different organizations adopting varying approaches to providing updates. For example, Oracle demonstrated consistent response times when releasing updates, while other vendors showed more variability. Additionally, the analysis revealed that community involvement was largely limited to identifying and reporting vulnerabilities, rather than offering solutions. However, this involvement can help speed up the CVE update process, as user-reported vulnerabilities tend to be resolved more quickly than those discovered solely by vendors. Vendors provided most fixes, and the time taken to resolve issues varied significantly.

We compiled a set of actionable insights to enhance the role of the community and improve monitoring practices for vulnerability management. First, we find that vendors should encourage community involvement beyond just issue reporting. This can improve response times, as community members might contribute patches or mitigations that complement vendor fixes. Second, software developers can adopt a multi-source monitoring approach by regularly checking platforms like GitHub, mailing lists, and public advisories for updates. Given that some fixes are released on one platform ahead of others, this approach can help developers respond quickly to vulnerabilities, reducing potential risks in dependent applications.

We highlight our contributions in this paper as follows:

\begin{itemize}
    \item We compile a comprehensive dataset that tracks software vulnerability response time, providing valuable insights for real-world practices. While previous datasets primarily analyze CVEs based on severity scores and other attributes, our dataset offers a comprehensive view of the entire vulnerability lifecycle, with a particular focus on Java vulnerabilities. This contribution enables further research in vulnerability management.
    
    \item We conduct an exploratory manual examination of CVE records and reference links, the study reveals key trends, such as vendors predominantly releasing fixes before announcing vulnerabilities and the community’s role being largely limited to vulnerability identification rather than providing direct fixes.

    \item We examine the relationship among various factors related to vulnerability and response handling, and we assist the developers in identifying the software library resident in the most responsive community.

    \item We offer practical recommendations, including encouraging greater community involvement in vulnerability remediation and promoting multi-source monitoring practices, which can help developers and organizations respond more quickly to vulnerabilities and reduce security risks associated with third-party libraries.
\end{itemize}

\textbf{Paper organization.} Section~\ref{sec:background} introduces the background of the vulnerability lifecycle and the motivation of our study. Section~\ref{sec:datasrc} describes the manual study setup. Section~\ref{sec:results} presents the results. Sections~\ref{sec:threats} and~\ref{sec:relatedworks} discuss threats to validity and related work. Section~\ref{sec:conclusions} concludes the paper.

\section{Background and Motivating Examples}
\label{sec:background}

In this section, we begin by outlining the context of vulnerability management. Following this, we present an illustrative example that motivates this study.

\subsection{Reporting and Managing Common Vulnerabilities and Exposures}

\phead{Common Vulnerabilities and Exposures (CVE) Program} is managed by the non-profit MITRE Corporation to maintain a publicly available list of recognized computer security vulnerabilities. Each entry, known as a \textbf{CVE Record}, includes a unique identifier (i.e., \textbf{CVE ID}, such as ``CVE-2014-12345''), a brief description of the vulnerability, and relevant references. In addition to MITRE, \textbf{CVE Numbering Authorities (CNAs)} are authorized to assign CVE IDs within their domains. CNAs comprise software vendors, open-source projects, and research groups, with over 400 organizations currently participating, including Red Hat, IBM, and Microsoft~\cite{CNA-partners}.

The \textbf{National Vulnerability Database (NVD)} offers supplementary information for each CVE Record, including severity scores based on the Common Vulnerability Scoring System (CVSS). The CVSS score assesses vulnerabilities using three metric groups—base, temporal, and environmental—ranging from 0.0 to 10.0, where higher values indicate greater severity.

\begin{figure}
    \centering
\includegraphics[width=1\linewidth]{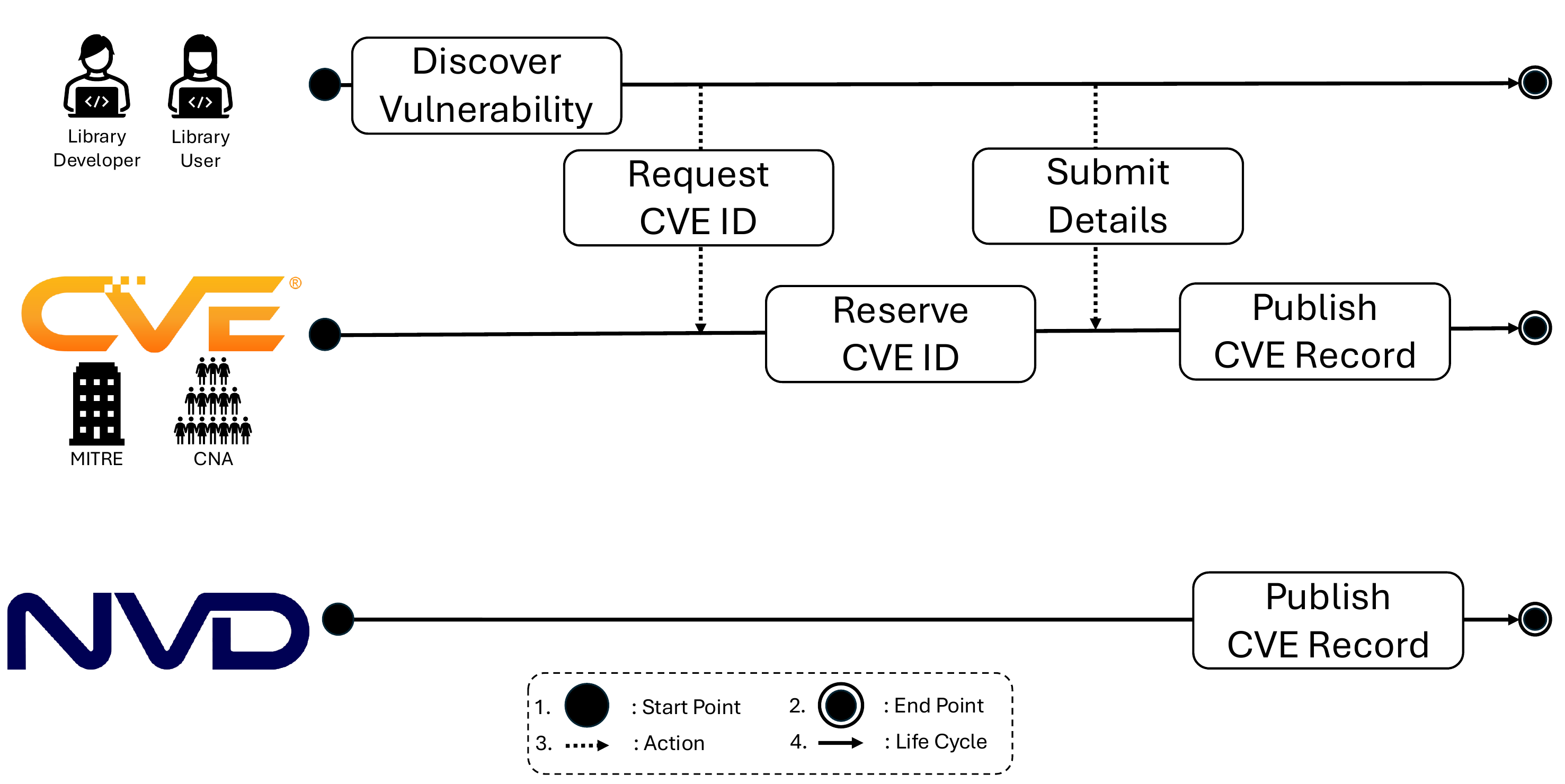}
  \caption{The lifecycle of how a CVE is discovered and recorded by developers, CNAs, MITRE, and NVD.} 
  \label{figure:CVELifecycle}
  \vspace{-0.3cm}
\end{figure}

\begin{figure}
  \centering
\subfloat[Empty CVE Record on December 4th, 2021\label{fig:empty_cve}]{%
\includegraphics[width=0.8\linewidth]{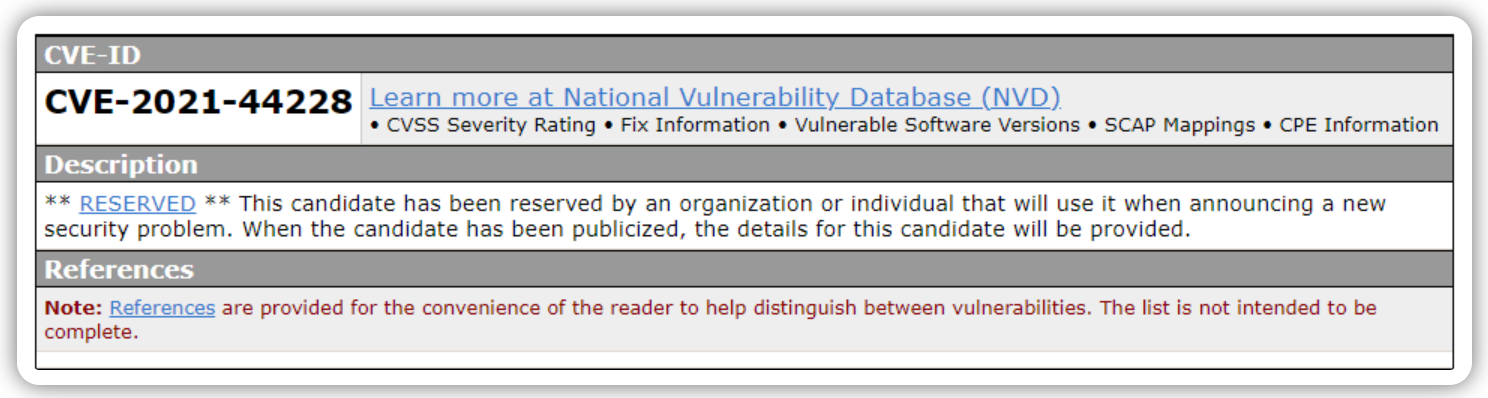}
    }
 \quad
\subfloat[Updated CVE Record on December 10th, 2021\label{fig:updated_cve}]{%
    \includegraphics[width=0.8\linewidth]{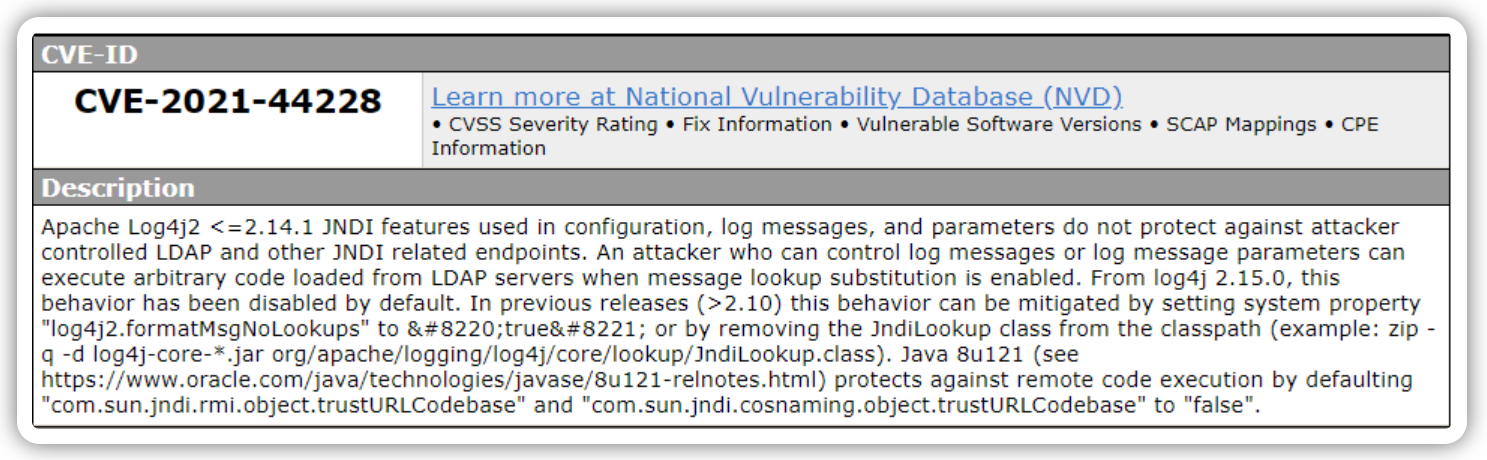}
    }
\caption{CVE-2021-44228 Record Changes: Initially reserved on December 4, 2021, with no details, the entry was updated on December 10, 2021, to include the root cause and potential mitigations.}
\label{fig:CVE_difference}
\end{figure}

\phead{Lifecycle of CVE} offers insights into how vulnerabilities are discovered, documented, and remediated, showing the dynamic nature of vulnerability management. Figure~\ref{figure:CVELifecycle} illustrates the complete lifecycle of a CVE. When an individual or organization discovers a vulnerability in a software library, they typically report it to the vendor. In some cases, the vendor may identify the vulnerability independently. Regardless of the source, both the library user and the vendor can contact MITRE or a CNA to request a CVE ID. MITRE then evaluates the vulnerability's severity and reserves the CVE ID, as shown in Figure~\ref{fig:empty_cve}. Once MITRE receives descriptive data about the vulnerability—including the names of affected products and versions, availability of fixes, vulnerability type, root cause, and impact~\cite{CNARules46:online}—the CVE Record is deemed complete. This record is then published on the CVE website, as seen in Figure~\ref{fig:updated_cve}, and added to the CVE List, making it accessible in the National Vulnerability Database.

\subsection{A Motivation Example}

In practice, the processes for discovering and addressing vulnerabilities can vary significantly. When developers identify a vulnerability, they typically either resolve it immediately or report it to the relevant vendor for remediation. This initial discovery represents a critical juncture in the vulnerability lifecycle, potentially leading to an immediate fix or necessitating additional time for appropriate mitigation.

The timeline from a vulnerability's initial discovery to its official publication on platforms such as the CVE and GitHub can vary widely. This variability is influenced by factors including the complexity of the required mitigation and the level of coordination among stakeholders.

Timely publication of vulnerabilities and their associated fixes is essential for developers, as it helps minimize security risks and enhances awareness within the software community. Therefore, understanding how vulnerabilities are managed throughout their lifecycle is crucial for software users, allowing them to effectively plan their updates and reduce potential exposure to threats.

We present the disclosure of Log4J vulnerabilities in November 2021 as a motivating example. 

\phead{(1) On December 10, 2021}, Apache released a security advisory~\cite{apache-log4j} for a critical remote code execution vulnerability in Log4j, affecting versions 2.0-beta9 to 2.15.0. This vulnerability, with a Base Score of 10.0—the maximum possible score—in the Common Vulnerability Scoring System~\cite{severityscore}, allowed unauthenticated attackers to execute arbitrary code. Given Log4j's extensive use in frameworks such as Spring, Netty, and MyBatis, it became a prime target for exploitation. In response, cybersecurity agencies in the UK, Australia, New Zealand, the US, and Canada issued alerts to raise awareness.

Tracked as CVE-2021-44228, this vulnerability was noted in a CVE created on November 26, 2021, but remained without description until the advisory was published, indicating that Apache was aware of the issue for two weeks prior to the fix.

\phead{(2) On December 13, 2021}, Log4j version 2.16.0 was released in Maven Central. This rapid update was atypical, as prior to 2021, Apache typically released fixes every 82 days~\cite{log4jreleasedate}. The following day, Apache announced CVE-2021-45046, indicating that version 2.15.0 was incomplete in certain configurations, with version 2.16.0 addressing this flaw. Notably, this CVE was created after the fix was released.

\phead{(3) On December 27, 2021}, Apache released Log4j versions 2.17.1, 2.12.4, and 2.3.2 to address CVE-2021-44832, which had a severity score of 6.6 and could allow attackers to modify configuration files for remote code execution. The CVE for this issue was created on December 11, 2021, but its description was not updated until after the fix was released. This suggests that while addressing CVE-2021-44228, Apache prioritized the more severe vulnerabilities.

Understanding the lifecycle of vulnerability management is crucial, as illustrated by the Log4j case. This example reveals different patterns in lifecycles: some vulnerabilities, such as CVE-2021-45046, had immediate fixes upon creation, while others, like CVE-2021-44228 and CVE-2021-44832, were created without immediate fixes or detailed information. This inconsistency can create an information gap for developers, leaving some aware of risks while others lack guidance on potential impacts and remediation steps. A better understanding of these lifecycle patterns could enable the community to develop more effective vulnerability management practices, ultimately reducing the time of exposure to security risks.

\section{Methodology}
\label{sec:datasrc}

\subsection{Data Collection}
\phead{Dataset.}
To conduct this study, we compiled a dataset of CVEs related to the Java ecosystem, given its broad use across software platforms and by major open-source foundations like Apache and Oracle. We began by downloading CVEs from the CVE database~\cite{CVE_MITRE} and filtering for records with the keyword \texttt{Java} resulting in 2,430 CVEs (data collected in January 2022). We then excluded records with \texttt{Javascript}, leaving 2,365 relevant CVEs. Finally, we focused on CVEs reported after 2017, narrowing the dataset to 735 records.

\subsection{Manual Study Process} 
\label{sec:manualstudy}
\phead{Preparation for Manual Analysis.}
To facilitate a comprehensive yet feasible analysis, we followed prior studies~\cite{huo2023autolog,li2023they} and employed random sampling to study a statistically significant sample of 312 CVEs, achieving a 98\% confidence level with a 5\% confidence interval~\cite{boslaugh2012statistics}.
For the 312 CVEs, we first collected the CVE reserved date, release date, description, vendor name, and severity score from the CVE and NVD websites. In contrast to prior studies~\cite{RUOHONEN2018239, 256930} that primarily rely on CVE and NVD data as well as selected mailing lists, we expanded our information collection to include reference links in the CVE and NVD reports, which encompass sources such as GitHub, ticketing platform, and relevant forums. 

We meticulously reviewed each link, focusing on key actions in the vulnerability lifecycle: \circled{1} Vulnerability Discovery, \circled{2} Fix Release, and \circled{3} Vulnerability Disclosure. Additionally, we analyzed the affected software versions and sought the release dates of fixes where applicable to enhance our understanding of vulnerability management.

\phead{Snowball Analysis.}
In addition to the reference links provided in the CVE reports, we employed a snowball analysis method~\cite{snowball} to explore all relevant links and their references. We analyzed each link recursively to construct a detailed timeline of each vulnerability's lifecycle. 
The first author conducted the data collection and analysis, while the second author verified labels and resolved disagreements. This highly labor intensive process took over four months and required approximately 1,000 man-hours. 
In total, we manually examined over 3,000 links derived through this snowball approach for a comprehensive understanding of the vulnerabilities and their context within the software ecosystem.

\subsection{Methodological Challenges and Mitigation Strategies} 
We further discuss the methodological challenges encountered during our analysis of CVE records, and outline the strategies we used to address these issues.

\phead{Fluctuating Data. }In our analysis of the CVEs related to Log4J, we encountered instances where a CVE is reserved both before and after a patch is released, and where the CVE description may initially be empty but is updated later. Vulnerabilities often undergo reanalysis when new information becomes available as vendors deprivatize certain data. To address these challenges, we utilize the Wayback Machine\footnote{\url{https://web.archive.org/}}, which enables access to archived versions of websites for the investigations of historical snapshots. By manually comparing different versions, we can identify changes in the description field and assess when specific events took place.

\phead{Uncertainties in Date and Time.} Time is crucial in software development, as it necessitates meticulous tracking of each event down to the second. However, some advisories present only the date without accompanying time specifications, complicating the reconstruction of event sequences. For instance, the Maven release date for the fix of CVE-2020-15777 is July 27, 2020~\cite{maven-threats-1}, which is the same date listed in the official advisory documentation~\cite{gradle-threats-1}. This lack of temporal detail can obscure the chronological order of actions taken. Vendors typically indicate in their announcements when certain fixes were unavailable at the time of publication~\cite{announce-before-release}. This allows us to assume that a fix is released prior to an advisory's publication unless explicitly stated otherwise.

\begin{figure}
    \centering
\includegraphics[width=1\linewidth]{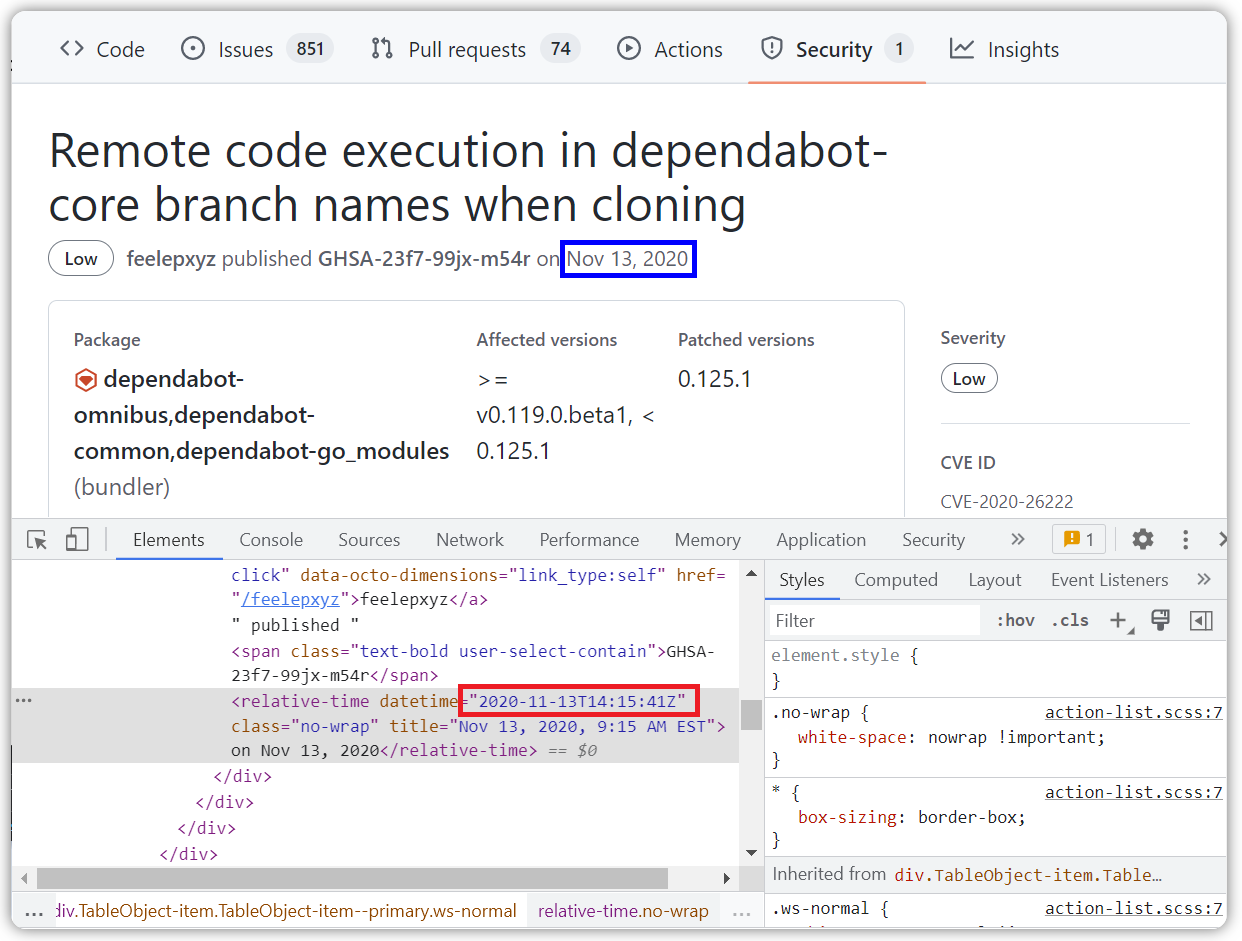}
  \caption{GitHub Security Advisory for CVE-2020-26222 with Inspect Element tool to unveil the specific timestamp of a change. } 
  \label{figure:InternalValidityTimeZone}
  \vspace{-0.3cm}
\end{figure}

To enhance accuracy, we extract timestamps from the analyzed websites, prioritizing the acquisition of ISO date-time format due to the variability in date and time presentations across different platforms. This is often facilitated by utilizing the Inspect Element feature within web browsers to access the underlying HTML code as seen in Figure~\ref{figure:InternalValidityTimeZone}. 
We convert all collected timestamps to a standardized time zone, such as UTC. If no time zone is specified, we default to UTC for consistency. Once standardized, events are organized chronologically, allowing for a coherent timeline reconstruction. This approach reduces discrepancies from varying time zone representations, enhancing our understanding of vulnerability management. If two events have the same date, we cross-check which event refers to the other. For example, if the disclosure date and fix release occur on the same day, we verify whether the disclosure references the fix and arrange the disclosure event before the fix in the timeline.

\phead{Issues with Rolling Update Dates.} Some vendors do not provide specific pages or links with publication dates for each vulnerability. Instead, updates are often appended to a single webpage or consolidated based on a time period. This practice can lead to inaccuracies when reconstructing the timeline of events. To address this, we utilize the Wayback Machine to compare content changes over time. However, timestamps retrieved from the Wayback Machine may be inaccurate due to potential delays in the web crawling process. 

To mitigate this issue, we cross-reference Wayback Machine data with other available sources, such as security advisories and release notes, to piece together a clearer picture of the vulnerability timeline.

\subsection{An Example of the Vulnerability Lifecycle} 
  \begin{table*}
  \centering
  \scriptsize
 \caption{{Events associated with CVE-2017-12628.}}
 \bgroup
 \def\arraystretch{1.12}
  \begin{tabular}{@{}r | p{7cm} | l@{}} 
\toprule
 \textbf{Event Date Time} & \textbf{Event Description} & \textbf{Event Category} \\ 
 \midrule
2017-08-06 20:00:00 & [CVE] Empty CVE Created & CVE Reserved \\
\textbf{2017-10-19 00:00:00} & \textbf{[Maven] Release Fix org.apache.james $>$ james-server-util $>$ 3.0.1} & \textbf{Vendor Provides Fix} \\
\textbf{2017-10-19 23:14:32} & \textbf{[Apache] Announce: Apache James 3.0.1 security release} & \textbf{Vendor Disclose} \\
2017-10-20 00:00:00 & [Apache] Security release: Apache James server 3.0.1 & Vendor Disclose \\
2017-10-20 05:57:59 & [GitHub] Release of james-project-3.0.1  & Vendor Provides Fix \\
2017-10-20 11:00:00 & [CVE] CVE Description updated & CVE Published \\
 \bottomrule
\end{tabular}
\egroup
\label{table:FlowSample}
\end{table*}

Table~\ref{table:FlowSample} illustrates the lifecycle associated with CVE-2017-12628. It presents the specific timestamps and the description of each event, along with manually assigned event categories. The lifecycle shows that the CVE ID was assigned on August 5th 2017 at 8pm, but no additional information was available at that time. On 19th October, Apache James 3.0.1 is available on Maven Repository. On the same day, Apache released an announcement about the new Apache James Server 3.0.1. One day later, Apache had another security release. This indicates that multiple related events can occur during a vulnerability's lifecycle; for example, fixes may be released on various platforms, and vendor statements may appear across different channels with intervals ranging from a few hours to several days.

To reduce the window of opportunity for attackers to exploit this vulnerability, our primary focus is on identifying the first significant event. We aim to understand the behaviors of software library vendors to enhance the immediate effectiveness of response efforts directed at addressing vulnerabilities.
\section{STUDY RESULTS}
\label{sec:results}

In this section, we discuss the results of our study by answering three research questions. 

\subsection{RQ1: What are the lifecycles of CVE reports?}
\phead{Motivation.} In this research question, we explore the lifecycles of CVE reports, specifically examining the flow of events and time gaps between the events that unfold from the initial discovery of a vulnerability to its public disclosure as a CVE report. Existing studies often overlook a comprehensive exploration of these lifecycles, leaving a gap in understanding how vulnerabilities are managed over time. By investigating this process, we aim to uncover the various actions taken by vendors in response to identified vulnerabilities, as well as the policies and practices they implement during this critical period. Such insights can assist developers in better planning their projects, ensuring they implement necessary updates and security measures in a timely manner.

\begin{center}
  \begin{table}
  \caption{Frequency of Identified Lifecycle timelines.}
\resizebox{1.0\columnwidth}{!}{%
\begin{tabular}{@{}l l@{}}  
 \toprule
 \textbf{Flow of events} & \textbf{Frequency (\%)} \\
 \midrule

 \textbf{Fix Available} & \textbf{305 (97.75\%)} \\ 

 \SFviii \textbf{Fix $\rightarrow$ Disclose} & \SFviii \textbf{268 (85.89\%)}  \\

 \SFxi~\SFviii Vendor Fix $\rightarrow$ Vendor Disclose & \SFxi~\SFviii 242 (77.56\%) \\
 
 \SFxi~\SFviii Vendor Fix $\rightarrow$ Community Disclose & \SFxi~\SFviii 24 (7.69\%) \\
 
 \SFxi~\SFviii Community Fix $\rightarrow$ Vendor Disclose & \SFxi~\SFviii 0 (0.00\%) \\
 
 \SFxi~\SFii Community Fix $\rightarrow$ Community Disclose & \SFxi~\SFii 2 (0.64\%) \\  
 
 \SFviii \textbf{Disclose $\rightarrow$ Fix} & \SFviii \textbf{19 (6.09\%)}  \\
 
 \SFxi~\SFviii Vendor Disclose $\rightarrow$ Vendor Fix & \SFxi~\SFviii 7 (2.24\%) \\
 
 \SFxi~\SFviii Vendor Disclose $\rightarrow$ Community Fix & \SFxi~\SFviii  1 (0.32\%) \\  
 
\SFxi~\SFviii Community Disclose $\rightarrow$ Vendor Fix & \SFxi~\SFviii  10 (3.21\%)  \\
 
 \SFxi~\SFii Community Disclose $\rightarrow$ Community Fix & \SFxi~\SFii 1 (0.32\%) \\ 
  
 \SFii \textbf{No Disclose} & \SFii \textbf{18 (5.77\%)}  \\

    ~\SFii Vendor Fix & ~\SFii 18 (5.77\%) \\

 \textbf{No Fix} & \textbf{7 (2.24\%)} \\ 
    ~\SFii Community Disclose & ~\SFii 7 (2.24\%)  \\
\bottomrule
 \end{tabular} 
 }
\label{tab:timelineresult}
\end{table}
\end{center}

\phead{Results.}

Table~\ref{tab:timelineresult} outlines the distribution of software library vulnerability lifecycle event flows. Our analysis reveals that 92.24\% of the 312 studied CVEs had an available fix, with the remainder being disclosed by the community without effective fixes being released. This suggests that the vendor is the primary entity responsible for addressing vulnerabilities, as fixes are more common than disclosures without fixes.

The most common lifecycle event flow, 77.56\%, shows that vulnerabilities were disclosed to the public only after the vendor’s fix was made available. Among these, 94.87\% of the disclosures were made by the vendor, with the remaining disclosures occurring via the community. \textbf{\textit{This highlights that vendors typically take a proactive role in managing vulnerabilities, prioritizing fixes before disclosing them to mitigate the risk of exploitation.}} In contrast, 2 cases are fixed and disclosed by the community and 10 cases showed that the community disclosed the vulnerability before notifying the vendor. \textbf{\textit{This indicates that community developers primarily act as whistleblowers rather than problem-solvers in the vulnerability lifecycle}}, often stepping in when vendors are slow to respond or fail to act within the given notification period. 

In 7.55\% of cases, the vulnerability was disclosed before the fix was released. 88.18\% of these disclosures were made by the vendor after the fix had already been released. \textbf{\textit{This suggests that vendors are aware of the risks associated with disclosure before a fix is available and prefer to release fixes before making vulnerabilities public}}. The vendor’s control over the timing of disclosure is crucial in reducing potential exploitation while ensuring that users are protected by fixes.

5.77\% of the vulnerabilities (18 cases) were initially disclosed by the community. Among these, 8 received a fix from the vendor, 3 were fixed by the community, and 7 did not receive any fix. While the community plays a crucial role in identifying and disclosing vulnerabilities, the responsibility for providing fixes often falls to the vendor. The presence of 7 cases without any fix suggests that \textbf{\textit{not all vulnerabilities lead to solutions, indicating gaps in the process where vulnerabilities remain unresolved despite disclosure}}. 

\finding{1}{Most vulnerabilities are fixed before being disclosed to the public. Vendors fixed most vulnerabilities, and community developers mainly act as whistleblowers in the lifecycle of vulnerabilities.}

\subsection{RQ2: What are the time gaps between different stages in the lifecycle of vulnerability management?}

\phead{Motivation.} Every minute a vulnerability remains unpatched, companies face the increasing risk of a Zero-Day Attack—a cyberattack exploiting a vulnerability discovered by attackers before the vendor is aware of it. Without a patch, these attacks are highly likely to succeed. For example, during the Apache Log4J incident, the release of the fix and public disclosure for CVE-2021-44228 occurred just hours apart, while for CVE-2021-45046, the security advisory was issued one day after the fix was available. 
The inconsistent handling of software vulnerabilities by vendors creates uncertainty for developers. When a vulnerability is discovered, or a new software version is released, developers must decide whether to update immediately or switch to an unaffected library. 

Understanding the time gaps between different stages of vulnerability management can help developers better anticipate when updates or fixes will be available. The findings can also provide insights into what kinds of preventative actions can be taken. Thus, in this RQ, we analyze the time gap between different stages in the lifecycles. 

\begin{figure*}
    \centering
    \includegraphics[width=0.8\linewidth]{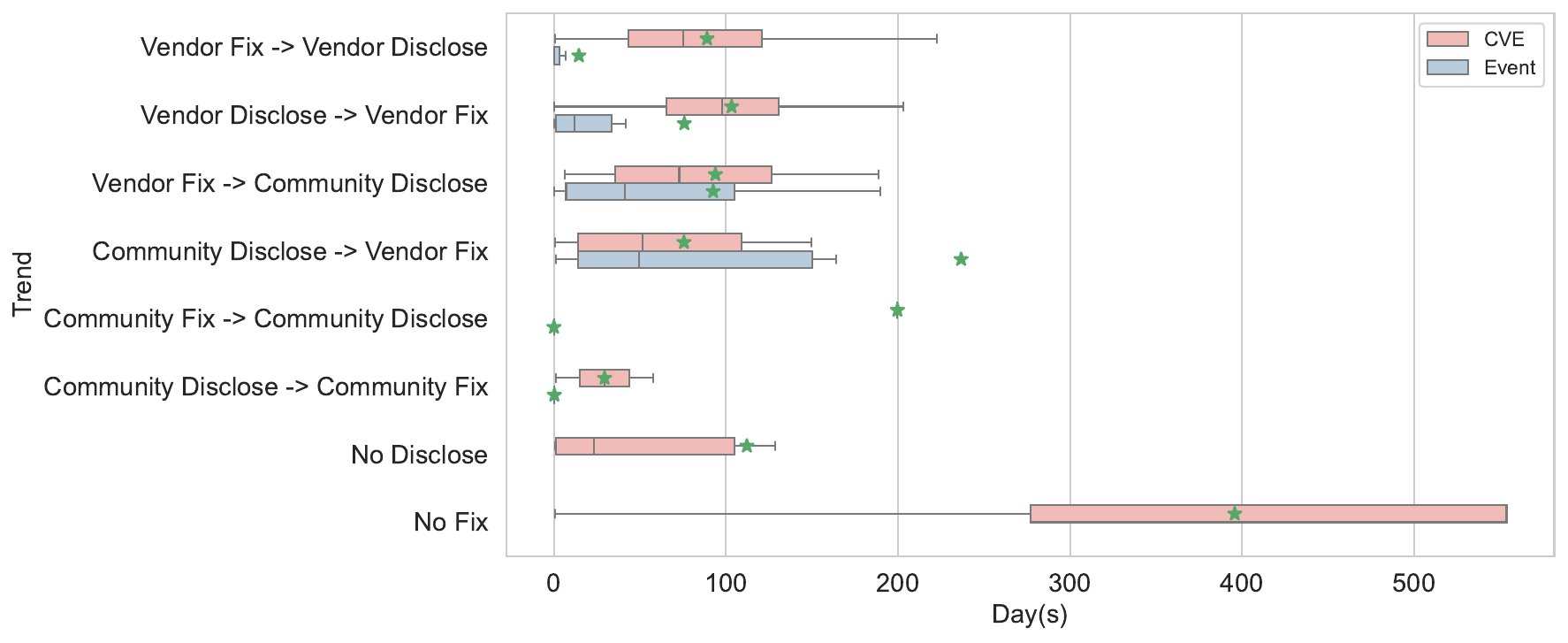}
    \caption{Number of days taken between events. Green stars indicate the mean values. } 
    \label{figure:days_by_trend}
\end{figure*}

\phead{Results.} Figure~\ref{figure:days_by_trend} illustrates the number of days taken between the actions. The data shows that vendor-driven processes for vulnerability management are the most common, with trends such as Vendor Fix $\rightarrow$ Vendor Disclose (242 occurrences, 77.56\%) and Vendor Disclose $\rightarrow$ Vendor Fix (7 occurrences, 2.24\%) making up 79.80\% of the cases. However, these vendor-driven processes tend to have longer response times compared to community-driven ones. For instance, the mean time for Vendor Fix $\rightarrow$ Vendor Disclose is 16.63 days, and 39.96 days for Vendor Disclose $\rightarrow$ Vendor Fix. In contrast, community-driven processes (i.e., Community Disclose $\rightarrow$ Community Fix and Community Fix $\rightarrow$ Community Disclose) are less frequent but much faster. Community Disclose $\rightarrow$ Community Fix occurred only 1 time (0.32\%), with a much shorter mean time of 0.47 days, much shorter than the vendor-driven timelines (e.g., 16.63 days for Vendor Fix $\rightarrow$ Vendor Disclose). Although these trends only account for a small portion of the data, community-driven processes like Community Fix $\rightarrow$ Community Disclose (with just 2 occurrences, 0.64\%) also show quicker resolution times, suggesting that community-driven vulnerability management may lead to faster outcomes.

We observe that 42 of the vendor announcements~\cite{notifyvendor1,notifyvendor2} 
feature a clear disclosure timeline, which includes timestamp of activities such as ``Vendor notification'' and ``Public disclosure''. In these cases, developers who identify vulnerabilities typically reach out to the affected organization to confirm the validity of their findings. Once the vulnerability is verified, both parties agree on a timeline for public disclosure, allowing the developer to share their findings only after the vendor has released a fix. The developer then documents the entire process, detailing the steps taken to uncover the vulnerability, and shares this information through various platforms, such as third-party blogs, GitHub Security Advisories, or their own blogs, along with a timeline of the discovery and correspondence. 

\begin{table}
    \centering
    \caption{Statistics on the Time Taken for CVE Updates: Comparison Between Cases with and Without User Notifying Vendors}
    \begin{minipage}[t]{0.45\textwidth}
        \centering
        \caption*{(a) With user notifying vendor (42 cases)}
        \begin{tabular}{l|c}
            \toprule
            Statistic & Value (days) \\
            \midrule
            Min & 0.0 \\
            1\textsuperscript{st} Quartile & 12.0 \\
            Median & 60.7 \\
            3\textsuperscript{rd} Quartile & 151.9 \\
            Max & 407.7 \\
            Mean & 104.0 \\
            Standard Deviation ($\sigma$) & 118.5 \\
            \bottomrule
        \end{tabular}
    \end{minipage}%
    \hfill
    \begin{minipage}[t]{0.45\textwidth}
        \centering
        \caption*{(b) Without user notifying vendor (270 cases)}
        \begin{tabular}{l|c}
            \toprule
            Statistic & Value (days) \\
            \midrule
            Min & 0.5 \\
            1\textsuperscript{st} Quartile & 46.7 \\
            Median & 73.3 \\
            3\textsuperscript{rd} Quartile & 124.6 \\
            Max & 995.7 \\
            Mean & 98.4 \\
            Standard Deviation ($\sigma$) & 109.8 \\
            \bottomrule
        \end{tabular}
    \end{minipage}
    \label{tab:maven_github}
\end{table}

To examine whether notifying vendors affects the overall CVE process, we analyzed the time taken for updates in cases where vendors were notified versus when they were not. On average, it takes slightly longer for users to notify the vendor about a vulnerability (104.0 days) compared to when the vendor discovers the issue on their own (98.4 days). \textbf{\textit{This suggests that vendor discovery might be slightly quicker, possibly due to vendors having internal monitoring and processes that allow them to identify vulnerabilities faster than users.}} However, user notifications tend to result in faster resolution times, as indicated by a lower median (60.7 days vs 73.3 days) and first quartile (12.0 days vs 46.7 days), suggesting that when users notify vendors, the response is generally quicker.

\finding{2}{The CVE process is expedited when vulnerabilities are reported by users.}

\begin{table}
    \centering
    \caption{Statistics of Response Times for Fix Releases: Days Between Maven and GitHub Releases.}
    \begin{minipage}[t]{0.45\textwidth}
        \centering
        \caption*{(a) GitHub releases fix after Maven (11 cases)}
        \begin{tabular}{l|c}
            \toprule
            Statistic & Value (days) \\
            \midrule
            Min & 1.0 \\
            1\textsuperscript{st} Quartile & 1.0 \\
            Median & 4.0 \\
            3\textsuperscript{rd} Quartile & 5.0 \\
            Max & 71.0 \\
            Mean & 10.9 \\
            Standard Deviation ($\sigma$) & 22.7 \\
            \bottomrule
        \end{tabular}
    \end{minipage}%
    \hfill
    \begin{minipage}[t]{0.45\textwidth}
        \centering
        \caption*{(b) Maven releases fix after GitHub (25 cases)}
        \begin{tabular}{l|c}
            \toprule
            Statistic & Value (days) \\
            \midrule
            Min & 1.0 \\
            1\textsuperscript{st} Quartile & 2.0 \\
            Median & 4.0 \\
            3\textsuperscript{rd} Quartile & 15.5 \\
            Max & 62.0 \\
            Mean & 13.8 \\
            Standard Deviation ($\sigma$) & 19.8 \\
            \bottomrule
        \end{tabular}
    \end{minipage}
    \label{tab:maven_github}
\end{table}

During our manual study, we observed that certain vendors release their fixes on both Maven Central and GitHub, prompting an investigation into which platform provides the fix first. Specifically, we aimed to determine the average number of days it takes for each platform to release a fix after the other. For CVE IDs with fixes available on both platforms, we identified 9 cases where Maven released a fix before GitHub, and 27 cases where GitHub was first. As shown in Table~\ref{tab:maven_github}, the statistics on response times for fix releases indicate that both platforms have varying response times. Still, there are notable differences in their mean and standard deviation.

While GitHub offers quicker releases in some cases, it has a higher standard deviation (58.6 days) than Maven (22.7 days), meaning Maven is more consistent in the time it takes to deliver fixes. Overall, Maven releases fixes faster on average (10.9 days vs. 13.8 days for GitHub) and with greater consistency in delivering timely fixes. However, we found that more fixes were released on GitHub before Maven. \textbf{\textit{This implies that relying on just one platform for fixes may not be sufficient; it is advisable to monitor both platforms for timely updates.}}

\finding{3}{Maven is more recommendable than GitHub due to its faster average release time (13.5 days vs. 26.5 days) and greater consistency (lower standard deviation of 20.1 vs. 58.6 days). However, one should monitor multiple platforms to receive timely updates.}

\subsection{RQ3: What are the characteristics of vulnerability response handling?}

\textbf{Motivation.} The unpredictable handling of vulnerabilities, exemplified by the Log4J case, poses significant challenges for developers, especially when no fix is provided. For instance, CVE-2018-20580, a vulnerability with a severity score of 8.8, went unanswered for 67 days, forcing the researcher to disclose it publicly. This leaves software library users reliant on affected software artifacts without official patches. Our findings in RQ1 indicate that vendor responses are a key factor in achieving timely resolutions of vulnerabilities. Therefore, selecting the right vendor is critical, as it directly affects the speed and effectiveness of fix releases. This study examines how developers choose software libraries amidst varying vendor practices in vulnerability management.

\begin{figure*}
  \centering
\subfloat[Distribution of Vendors \label{fig:vendor_distribution}]{%
     \includegraphics[width=0.30\linewidth]{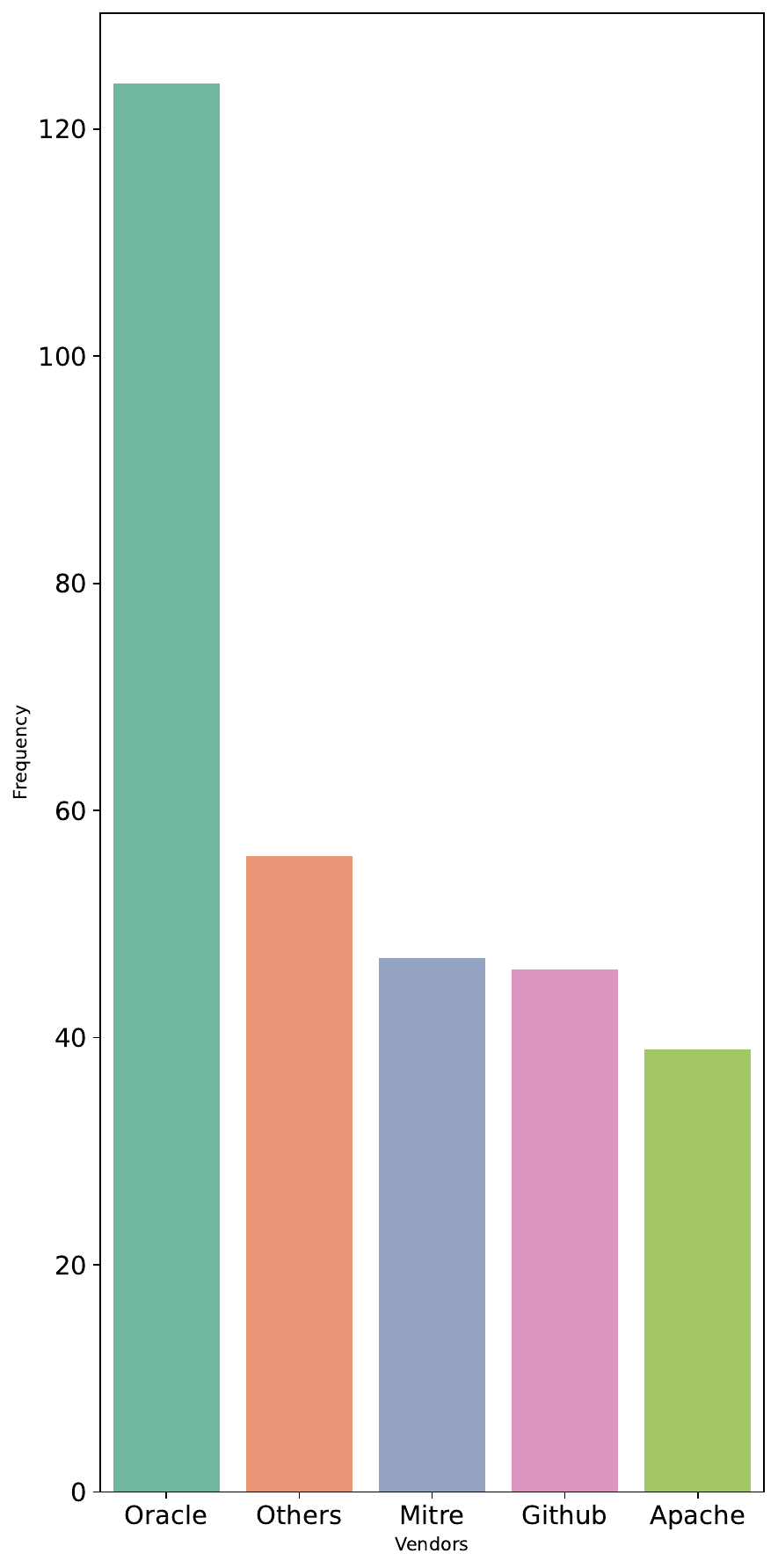}}
 \quad
\subfloat[Severity Score Analysis \label{fig:vendor_severity}]{%
      \includegraphics[width=0.30\linewidth]{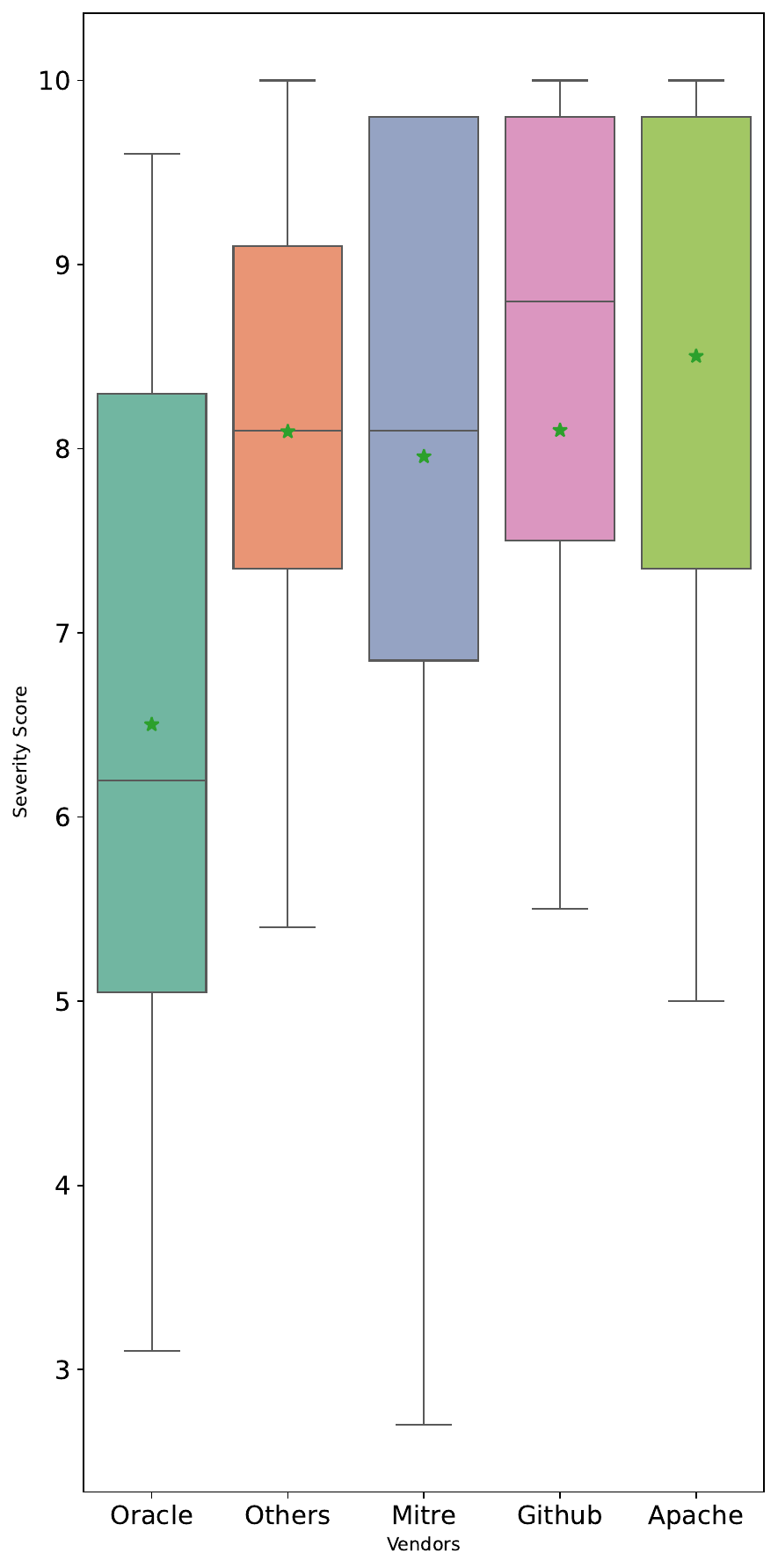}}
\subfloat[Number of Days to Update CVE \label{fig:vendor_updateCVE}]{%
     \includegraphics[width=0.30\linewidth]{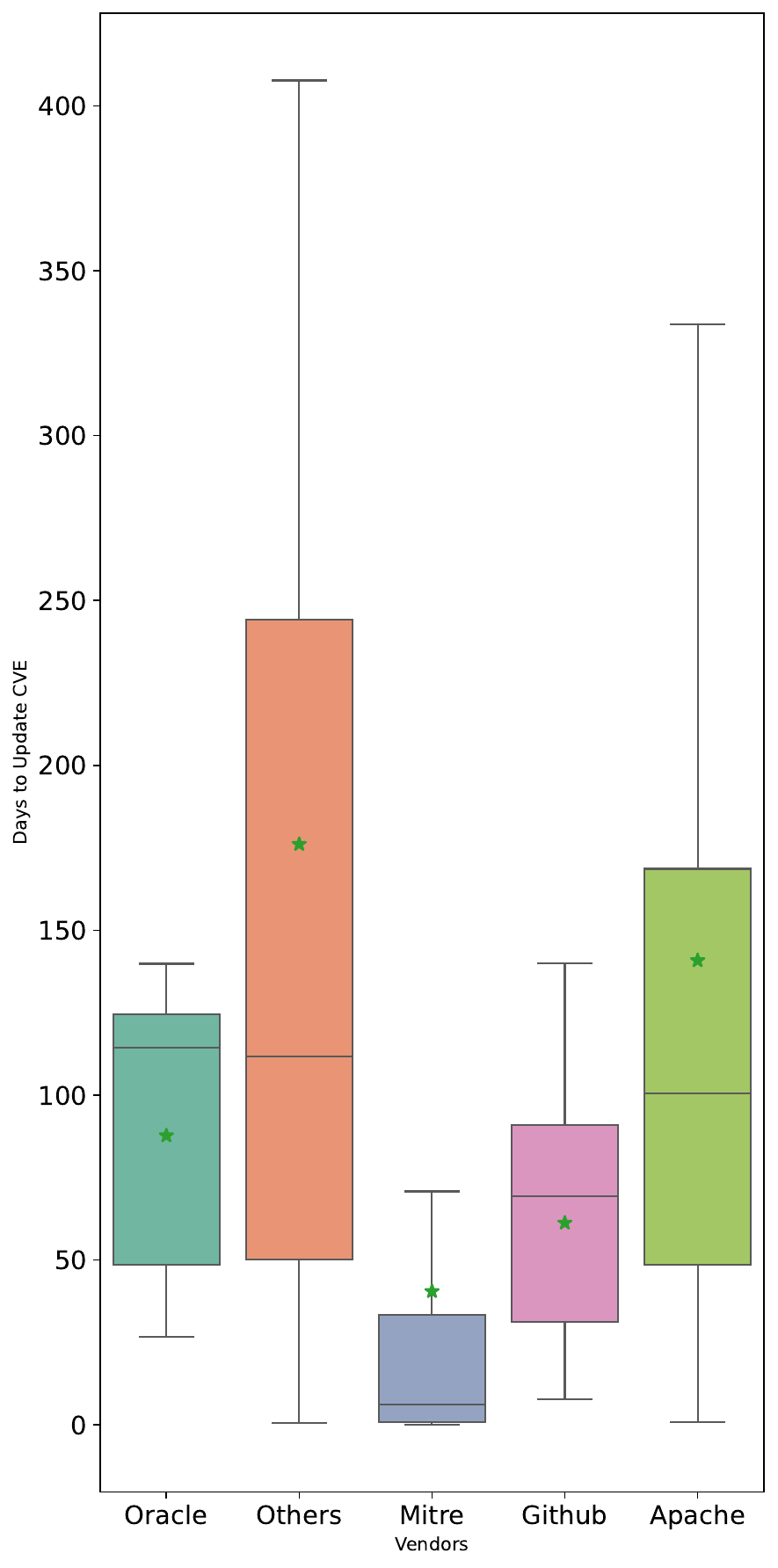}}
 \quad
\caption{Vendor Insights: A Comprehensive Analysis of Vendor Distribution, Severity Scores, and Update Timeliness. Green stars indicate the mean values.}
\label{fig:vendor}
\end{figure*}

\phead{Results.} 

Figure~\ref{fig:vendor} shows that Oracle has more vulnerabilities than other vendors (120 reported vulnerabilities, while every other vendor has less than 60). This suggests that vulnerabilities or issues associated with Oracle are more prevalent or reported more frequently in the dataset. Both Oracle and Apache are large organizations with numerous Java-based products, which may contribute to a higher likelihood of reported vulnerabilities~\cite{apacheprojects}. However, they differ in several key areas. Oracle has a higher frequency of vulnerabilities (124 cases vs. 39 cases by Apache) but lower severity on average, suggesting that while vulnerabilities are more commonly reported, they tend to be less critical. The high frequency for Oracle indicates that it may be more actively monitored or that vulnerabilities are more commonly reported. In contrast, Apache’s vulnerabilities tend to be more severe, with a mean severity score of 8.50 compared to Oracle's 6.50. Despite the higher severity, Apache’s response times are slower, with an average of 140.80 days and a maximum of 472.35 days to update CVEs, which could increase risk exposure for users relying on their products. 

\finding{4}{Oracle has a higher frequency of vulnerabilities (124 cases), with medium severity (mean score of 6.5), and a faster response time (average of 39.30 days to update CVEs). In contrast, Apache has fewer occurrences (39 cases), but those vulnerabilities are of higher severity (mean score of 8.5) and take longer to update (average of 118.98 days). }

When a source is attributed to MITRE, it indicates that the vulnerability was identified by researchers affiliated with MITRE. Similarly, vulnerabilities linked to GitHub are retrieved from Security Advisories hosted on the GitHub platform. While the information may originate from the original vendor who uploaded it to their GitHub account, these sources are not part of the Cybersecurity Notification Action (CNA). CNAs are organizations authorized by the CVE Program to assign CVE IDs and publish CVE Records, lending them a level of prestige and credibility. This aligns with our observation that MITRE has a mean severity score of 7.96 (moderate), indicating a reasonable level of severity, while GitHub’s mean score of 8.10 (high) suggests that its vulnerabilities tend to be more critical. "Others" has a mean of 176.07 days to update CVEs and a maximum of 995.66 days, highlighting potential challenges in managing vulnerabilities. \textbf{\textit{This indicates that these vendors may lack the same level of resources or response capabilities as larger companies.}}

\finding{5}{Compared to Apache and Oracle, MITRE and GitHub have higher severity scores (7.96 and 8.10, respectively) and faster CVE update times (40.47 and 61.23 days). }

We conducted a correlation analysis to explore the relationship between the events we derived and key metrics of the CVEs in our dataset, such as severity scores and severity categories. By calculating Pearson correlation coefficients~\cite{pearson} through a correlation matrix, we aimed to identify significant associations that could deepen our understanding of how these metrics interact and influence the timelines. A correlation value closer to -1 or 1 indicates strong negative and positive relationships, while values near 0 suggest no correlation. 
We find that the correlation values range from -0.12 to 0.07. The weak correlations indicate there is no significant relationship between the events and CVE metrics, \textbf{\textit{suggesting limited consistency in vulnerability management practices}}. 

\finding{6}{The Spearman correlations between the uncovered events, NIST, and CNA Severity Scores range from -0.12 to 0.07, indicating limited consistency in vulnerability management practices.}
\section{Discussions}
\label{sec:implications}

In this section, we highlight the actionable insights and potential future directions based on our findings.

\subsection{Implications for Software Library Developers}

\phead{Prioritize Timely Fix Releases and Disclosure. }
As highlighted in RQ1, while vendors are the primary source of fixes, they often experience delays in disclosing them. 
Apart from releasing fixes promptly once a vulnerability is found, it is also crucial to expedite the CVE update process to ensure that vulnerabilities are disclosed in a timely manner.

Delays in updating CVE records extend the vulnerability lifecycle, leaving systems exposed for longer. Major Java library providers like Oracle and Apache have shown significant delays, with median times of 114 days for Oracle and 100 days for Apache. On average, it takes 22.1 days for vulnerabilities to be fixed, but the time from Vendor Fix $\rightarrow$ Vendor Disclose can stretch up to 632 days in some cases, which highlights the urgent need to speed up the CVE update process. Vendors should focus on accelerating both the release of fixes and the update of CVE records to reduce exposure and enhance the overall security of the ecosystem.

\phead{Encourage Community Participation on Vulnerability Resolution. }
Our findings revealed that community developers primarily act as whistleblowers rather than problem-solvers in the vulnerability lifecycle. However, when community developers notify vendors about vulnerabilities, the median time to fix is 60.7 days, compared to 73.3 days when vendors discover the issue on their own. This suggests that user-reported vulnerabilities tend to be resolved faster. Proactive communication from users can accelerate the identification and resolution of security issues. Therefore, it is crucial to establish standardized communication protocols to ensure that vendors can efficiently receive and act on these reports.

\subsection{Implications for Software Library Users}

\phead{Encouraging Fix Monitoring During Vendor Disclosure Delays. }
In RQ1, we find that 96.47\% of the fixes are provided by vendors. However, as observed in RQ2, vendors take an average of 16.63 days to disclose a fix, and the process of updating the CVE typically takes an average of 91.99 days after the fix is implemented. This suggests that while vendors provide the fixes, there is often a delay in disclosing them, possibly to allow users time to apply fixes before the vulnerability becomes widely known. Nonetheless, vendors still need to communicate the significance of the fix to software library users. Given these delays, software library users should actively monitor available fixes and not rely solely on CVE updates or official vendor disclosures on CVE websites.

\phead{Enhancing Vulnerability Management through Multi-Source Monitoring. }
Software library users should adopt comprehensive monitoring practices by utilizing multiple channels -- such as public advisories, mailing lists, and code repositories -- to effectively detect and address vulnerabilities. Our findings show that in 245 out of 312 cases, vendors announced the release of a fix before it was officially recorded in the CVE database. This underscores the importance of monitoring various sources, including GitHub Security Advisories, Openwall (a third-party site), and the vendors' own websites, to obtain timely information about vulnerabilities.

As highlighted in RQ2, different companies have varying release practices: some publish fixes on Maven first, while others prefer GitHub. For instance, during the Apache Log4j incident, two patches~\cite{apachePatch, apachePatch2} with an ``rc1'' suffix was released one day prior to the official patch. This practice enables users to anticipate upcoming fixes and predict release timelines by actively monitoring both Maven and GitHub.

GitHub is a valuable resource for monitoring vulnerabilities, as tracking commit histories and issue trackers can reveal recent fixes and ongoing discussions. By keeping up with recent commits and following developer communications, users can gain insights into emerging issues and potential solutions before an official vendor release. This multi-source approach provides earlier warnings and enables organizations to proactively mitigate risks before formal patches become available.

\phead{Evaluate Vendor Reliability in Frequency, Severity, and Response Times.}
When evaluating vendors, it is important to consider both the frequency of vulnerabilities and their severity. High-frequency, lower-severity vulnerabilities may be less critical than those with lower frequency but higher severity. Additionally, choosing reliable vendors and evaluating their response times for addressing software issues is essential for safeguarding software environments. The variability in vendor response times suggests that users should prioritize libraries maintained by vendors with a proven track record of promptly addressing vulnerabilities. For example, Oracle has an average resolution time of 87 days, compared to Apache's 140 days. While Apache may occasionally provide quicker fixes, its higher mean and greater variability ($\sigma$ = 118.86) indicate less consistent response. In contrast, Oracle ($\sigma$ = 39.29) generally offers fixes in a more timely and reliable manner.

\phead{Automate Dependency Management and Updates.}
Automating dependency management and updates can be a valuable strategy for minimizing the risks associated with third-party libraries. Several tools (e.g., Dependabot by GitHub~\cite{dependabot}) automatically detect and apply updates for known vulnerabilities, which enable organizations to streamline the process of maintaining a secure software environment. Additionally, automating GitHub issue tracking of the software library in use can help software library users to stay alerted to potential vulnerabilities as they are raised and discussed by the developer community, further enabling quicker assessment and action. 

\section{Threats to Validity}
\label{sec:threats}
This section discusses the limitations and potential threats to the validity of our observations on the manual study.

\phead{Construct Validity. }
The construct validity of our study is fundamentally linked to our data collection methodology. First, we conducted the research on a statistically significant sample drawn from randomly selected CVEs, employing a 95\% confidence level and a 5\% margin of error, which may introduce minor noise. Second, we filtered the CVE reports using the keyword `Java'', which could limit the comprehensiveness of our findings by overlooking some seemingly trivial Java-related CVEs that contained minimal reported information. Third, interpretations of vulnerability reports and disclosure announcements can vary. To mitigate this issue, two authors of this paper independently reviewed the relevant reports and discussed any ambiguities until a consensus was reached.

\phead{Internal Validity. }
We observe that not all websites display publication times in the same time zone; nevertheless, we strive to convert the date and time to our local time zone whenever specified or when presented in a universal timestamp format. Time is crucial in software development, prompting us to track events with precision down to the second; however, some advisories only provide the date without the accompanying time. Vendors typically indicate in announcements when listed fixes were unavailable at the time of publication, leading us to assume that a fix is released prior to the announcement unless stated otherwise. Additionally, some vendors do not provide specific pages or links with publication timestamps for each vulnerability, opting instead to consolidate updates on a single webpage or group them by time periods, which can lead to inaccuracies in event reconstruction. To mitigate this, we use the Wayback Machine to compare content changes, although the timestamps obtained may be unreliable due to delays in the web crawling process. Furthermore, public disclosure policies and agreements between vendors and vulnerability discoverers can result in specific information being revealed later, potentially altering the sequence of events. To address these challenges, we focus on excluding newly discovered cases and ensure that we consider only the dates when information was publicly accessible during our reconstruction of events, allowing for a more accurate timeline.

\phead{External Validity. }
The relevance of this study may be challenged because it examines cases from the past five years, during which new software practices and policies could significantly alter crowd responses, leading to potential timeliness concerns. Furthermore, the findings are confined to the Java software ecosystem, suggesting that the observed trends and reactions may not be applicable to vulnerabilities in other platforms. Future research could investigate other dependency management systems to gain a more comprehensive understanding.

\section{Related Work}
\label{sec:relatedworks}

\phead{Vulnerability Lifecycle.}
MITRE researchers~\cite{Gertner_Zaromb_Schneider_Roberts_Matthews_2016} uncovered researchers' bias in selecting software and the methodology used to test the software for vulnerabilities. Researchers tend to analyze vulnerability statistics using large data repositories and add them to their database when the data fit their hypothesis. Therefore, we scour all kinds of platforms to prevent researchers' bias. Khanmohammadi et al.~\cite{Khanmohammadi2023HalfDayVA} based their study on CVE reports, and examined only the early days of CVE reports being modified and updated in the first days after their initial disclosure. 
Frei et al.~\cite{frei-similarstudies-1, frei-similarstudies-2} performed a comprehensive study of the life-cycle of vulnerabilities, ranging from discovery to patch. Similar to our study, these studies are based on extensive data mining on commit logs, bug trackers, and mailing lists. They determined the discovery, disclosure, exploit, and patch dates of 14,000 vulnerabilities between 1996 and 2006. This data is then used to plot different dates against each other, e.g. discovery date vs disclosure date, exploit date vs disclosure date, and patch date vs. disclosure date. Their study found that black hat hackers are fast in creating exploits for new vulnerabilities and that the number of zero-day attacks is increasing rapidly. Our study updates these findings as their data has been outdated. 
Sommestad et al.~\cite{effort_estimate} reviewed the vulnerability discovery aspect, and their results showed that Vulnerability Discovery Projects require 14 days of effort to discover a software vulnerability. However, their paper did not cover what happens after the vulnerability is exposed~\cite{mitre_bias}. Our work covers the complete timeline by including the study of the subsequent development of events.

\phead{Vulnerability Study based on Vulnerability Databases.}
Since its inception in 1997, NVD has published information about more than 43,000 software vulnerabilities affecting more than 17,000 software applications. However, data supplied by NVD has poor prediction capability~\cite{nvd-not-good-enough-example,nvd-not-good-enough-example2}. This claim is further supported by Zhang et al.\cite{nvd-not-good-enough-example2}, who use several machine learning algorithms to predict the Time To Next Vulnerability (TTNV) for various software applications. The authors claimed that the quality of the NVD is lacking and are only content with their predictions for a few vendors. Massacci et al.~\cite{right_source_vulnerability, vulnerability_prediction} also supported this claim further by performing a comparative breakdown of public vulnerability databases and compiling them into a joint database for analysis. A table of recent authors' usage of vulnerability databases in similar studies is collected, categorizing these studies as prediction, modelling, or fact-finding. They showed that by using two different data sources for conducting the same experiment, the results could be significantly different due to the high degree of inconsistency in the data available to the research community during the research period. They further tried to confirm the correctness of their database by comparing data from different sources. 
Ozment~\cite{ozment_vulnerability} resonated with the sentiment in studies above after examining the vulnerability discovery models proposed by Malaiya et al.~\cite{malaiya_prediction}. He identified some limitations that make these models inapplicable, citing insufficient information included in a government-supported vulnerability database, such as NVD, as one of the reasons. In our study, we collected information about the vulnerability from several sources, such as official websites, email correspondences, software artifact repositories, and issue-tracking platforms to avoid inaccurate results due to low-quality information.
\section{Conclusions}
\label{sec:conclusions}

In this paper, we conducted a manual analysis of Java-related CVEs to identify common trends in the timeline and community reactions to vulnerabilities. By examining a variety of sources — including official vendor documentation, developer emails, software repositories, and issue tracking platforms — we reconstructed the sequence of events for each vulnerability and created a comprehensive dataset that captures the detailed timeline of vulnerability handling. This allowed us to identify key trends in the handling and disclosure of vulnerabilities. We calculated the differences in reaction times across these trends and explored additional factors that may have influenced the variability. Overall, our study provides valuable insights into the real-world patterns of software vulnerability management, offering a clearer understanding of how the vendors and developer communities respond throughout the vulnerability lifecycle.


\balance
\bibliographystyle{IEEEtran}
\bibliography{paper}

\begin{thebibliography}{10}
\providecommand{\url}[1]{#1}
\csname url@samestyle\endcsname
\providecommand{\newblock}{\relax}
\providecommand{\bibinfo}[2]{#2}
\providecommand{\BIBentrySTDinterwordspacing}{\spaceskip=0pt\relax}
\providecommand{\BIBentryALTinterwordstretchfactor}{4}
\providecommand{\BIBentryALTinterwordspacing}{\spaceskip=\fontdimen2\font plus
\BIBentryALTinterwordstretchfactor\fontdimen3\font minus
  \fontdimen4\font\relax}
\providecommand{\BIBforeignlanguage}[2]{{%
\expandafter\ifx\csname l@#1\endcsname\relax
\typeout{** WARNING: IEEEtran.bst: No hyphenation pattern has been}%
\typeout{** loaded for the language `#1'. Using the pattern for}%
\typeout{** the default language instead.}%
\else
\language=\csname l@#1\endcsname
\fi
#2}}
\providecommand{\BIBdecl}{\relax}
\BIBdecl

\bibitem{kula}
R.~Kula, D.~German, A.~Ouni, T.~Ishio, and K.~Inoue, ``Do developers update
  their library dependencies?'' \emph{Empirical Software Engineering}, vol.~23,
  pp. 1--34, 02 2018.

\bibitem{wallace}
D.~Wallace and R.~Fujii, ``Software verification and validation: an overview,''
  \emph{IEEE Software}, vol.~6, no.~3, pp. 10--17, 1989.

\bibitem{cve-distribution}
``Nvd - cvss severity distribution over time,''
  \url{https://nvd.nist.gov/general/visualizations/vulnerability-visualizations/cvss-severity-distribution-over-time\#CVSSSeverityOverTime},
  [n.\,d.], (Accessed on 12/04/2022).

\bibitem{log4j-advisory-1}
``{GHSA-jfh8-c2jp-5v3q | GHSA | Open Source Insights},''
  \url{https://deps.dev/advisory/GHSA/GHSA-jfh8-c2jp-5v3q}, [n.\,d.], (Accessed
  on 12/04/2022).

\bibitem{log4j-advisory-2}
``{GHSA-7rjr-3q55-vv33 | GHSA | Open Source Insights},''
  \url{https://deps.dev/advisory/GHSA/GHSA-7rjr-3q55-vv33}, [n.\,d.], (Accessed
  on 12/04/2022).

\bibitem{google-log4j}
``Google online security blog: Understanding the impact of apache log4j
  vulnerability,''
  \url{https://security.googleblog.com/2021/12/understanding-impact-of-apache-log4j.html},
  [n.\,d.], (Accessed on 12/01/2022).

\bibitem{software-vulnerability-prediction}
``Software vulnerability prediction using text analysis techniques,''
  \url{https://dl.acm.org/doi/pdf/10.1145/2372225.2372230}, [n.\,d.], (Accessed
  on 12/04/2022).

\bibitem{software-vulnerability-prediction-2}
``Vulnerability prediction from source code using machine learning | ieee
  journals and magazine | ieee xplore,''
  \url{https://ieeexplore.ieee.org/abstract/document/9167194}, [n.\,d.],
  (Accessed on 12/04/2022).

\bibitem{TrendAnalysis}
S.~Neuhaus and T.~Zimmermann, ``Security trend analysis with cve topic
  models,'' in \emph{2010 IEEE 21st International Symposium on Software
  Reliability Engineering}, 2010, pp. 111--120.

\bibitem{Yosifova2021PredictingVT}
\BIBentryALTinterwordspacing
V.~Yosifova, A.~Tasheva, and R.~Trifonov, ``Predicting vulnerability type in
  common vulnerabilities and exposures (cve) database with machine learning
  classifiers,'' \emph{2021 12th National Conference with International
  Participation (ELECTRONICA)}, pp. 1--6, 2021. [Online]. Available:
  \url{https://api.semanticscholar.org/CorpusID:237246224}
\BIBentrySTDinterwordspacing

\bibitem{nvd-not-good-enough-example}
``An empirical study on using the national vulnerability database to predict
  software vulnerabilities | proceedings of the 22nd international conference
  on database and expert systems applications - volume part i,''
  \url{https://dl.acm.org/doi/10.5555/2035368.2035388}, [n.\,d.], (Accessed on
  12/04/2022).

\bibitem{nvd-not-good-enough-example2}
S.~Zhang, D.~Caragea, and X.~Ou, ``An empirical study on using the national
  vulnerability database to predict software vulnerabilities,'' in
  \emph{Database and Expert Systems Applications}, A.~Hameurlain, S.~W. Liddle,
  K.-D. Schewe, and X.~Zhou, Eds., 2011, pp. 217--231.

\bibitem{Li2023TheAO}
\BIBentryALTinterwordspacing
X.~Li, S.~Moreschini, Z.~Zhang, F.~Palomba, and D.~Taibi, ``The anatomy of a
  vulnerability database: A systematic mapping study,'' \emph{J. Syst. Softw.},
  vol. 201, p. 111679, 2023. [Online]. Available:
  \url{https://api.semanticscholar.org/CorpusID:264407972}
\BIBentrySTDinterwordspacing

\bibitem{CNA-partners}
``List of partners | cve,''
  \url{https://www.cve.org/PartnerInformation/ListofPartners}, (Accessed on
  10/30/2024).

\bibitem{CNARules46:online}
``Cna rules | cve,''
  \url{https://www.cve.org/ResourcesSupport/AllResources/CNARules\#section_8-1_cve_record_information_requirements},
  (Accessed on 01/17/2023).

\bibitem{apache-log4j}
``Log4j – apache log4j security vulnerabilities,''
  \url{https://logging.apache.org/log4j/2.x/security.html}, [n.\,d.], (Accessed
  on 11/29/2022).

\bibitem{severityscore}
``Nvd - vulnerability metrics,'' \url{https://nvd.nist.gov/vuln-metrics/cvss},
  (Accessed on 10/31/2024).

\bibitem{log4jreleasedate}
``Maven repository: org.apache.logging.log4j > log4j-core,''
  \url{https://mvnrepository.com/artifact/org.apache.logging.log4j/log4j-core},
  (Accessed on 10/31/2024).

\bibitem{CVE_MITRE}
\url{https://cve.mitre.org/data/downloads/index.html}, [Accessed 28-10-2024].

\bibitem{huo2023autolog}
Y.~Huo, Y.~Li, Y.~Su, P.~He, Z.~Xie, and M.~R. Lyu, ``Autolog: A log sequence
  synthesis framework for anomaly detection,'' in \emph{2023 38th IEEE/ACM
  International Conference on Automated Software Engineering (ASE)}.\hskip 1em
  plus 0.5em minus 0.4em\relax IEEE, 2023, pp. 497--509.

\bibitem{li2023they}
Z.~Li, A.~R. Chen, X.~Hu, X.~Xia, T.-H. Chen, and W.~Shang, ``Are they all
  good? studying practitioners' expectations on the readability of log
  messages,'' in \emph{2023 38th IEEE/ACM International Conference on Automated
  Software Engineering (ASE)}, 2023, pp. 129--140.

\bibitem{boslaugh2012statistics}
\BIBentryALTinterwordspacing
S.~Boslaugh, \emph{Statistics in a Nutshell, 2nd Edition}.\hskip 1em plus 0.5em
  minus 0.4em\relax O'Reilly Media, Incorporated, 2012. [Online]. Available:
  \url{https://books.google.ca/books?id=s1llAQAACAAJ}
\BIBentrySTDinterwordspacing

\bibitem{RUOHONEN2018239}
\BIBentryALTinterwordspacing
J.~Ruohonen, S.~Rauti, S.~Hyrynsalmi, and V.~Leppänen, ``A case study on
  software vulnerability coordination,'' \emph{Information and Software
  Technology}, vol. 103, pp. 239--257, 2018. [Online]. Available:
  \url{https://www.sciencedirect.com/science/article/pii/S0950584917305116}
\BIBentrySTDinterwordspacing

\bibitem{256930}
\BIBentryALTinterwordspacing
A.~D. Householder, J.~Chrabaszcz, T.~Novelly, D.~Warren, and J.~M. Spring,
  ``Historical analysis of exploit availability timelines,'' in \emph{13th
  USENIX Workshop on Cyber Security Experimentation and Test (CSET 20)}.\hskip
  1em plus 0.5em minus 0.4em\relax USENIX Association, Aug. 2020. [Online].
  Available:
  \url{https://www.usenix.org/conference/cset20/presentation/householder}
\BIBentrySTDinterwordspacing

\bibitem{snowball}
T.~Greenhalgh and R.~Peacock, ``Effectiveness and efficiency of search methods
  in systematic reviews of complex evidence: Audit of primary sources,''
  \emph{BMJ (Clinical research ed.)}, vol. 331, pp. 1064--5, 12 2005.

\bibitem{maven-threats-1}
``Maven repository: com.gradle > gradle-enterprise-maven-extension > 1.6,''
  \url{https://mvnrepository.com/artifact/com.gradle/gradle-enterprise-maven-extension/1.6},
  [n.\,d.], (Accessed on 12/04/2022).

\bibitem{gradle-threats-1}
``Gradle enterprise maven extension user manual | gradle enterprise docs,''
  \url{https://docs.gradle.com/enterprise/maven-extension/\#1_6}, [n.\,d.],
  (Accessed on 12/04/2022).

\bibitem{announce-before-release}
``[security] cve-2019-0232 apache tomcat remote code execution on
  windows-apache mail archives,''
  \url{https://lists.apache.org/thread/8r4sf1xz8kxl2k2pg43g45ghorcfbxrq},
  [n.\,d.], (Accessed on 12/04/2022).

\bibitem{notifyvendor1}
``Full disclosure: Open-xchange security advisory 2021-11-18,''
  \url{https://seclists.org/fulldisclosure/2021/Nov/42}, 10 2024, (Accessed on
  10/30/2024).

\bibitem{notifyvendor2}
``advisories/atredis-2019-0006.md at master · atredispartners/advisories ·
  github,''
  \url{https://github.com/atredispartners/advisories/blob/master/ATREDIS-2019-0006.md},
  10 2024, (Accessed on 10/30/2024).

\bibitem{apacheprojects}
``Apache projects list,'' \url{https://projects.apache.org/projects.html},
  (Accessed on 10/30/2024).

\bibitem{pearson}
J.~Benesty, J.~Chen, Y.~Huang, and I.~Cohen, ``Pearson correlation
  coefficient,'' in \emph{Noise reduction in speech processing}.\hskip 1em plus
  0.5em minus 0.4em\relax Springer, 2009, pp. 37--40.

\bibitem{apachePatch}
``Release log4j-2.12.4-rc1 · apache/logging-log4j2,''
  \url{https://github.com/apache/logging-log4j2/releases/tag/log4j-2.12.4-rc1},
  (Accessed on 10/30/2024).

\bibitem{apachePatch2}
``Release log4j-2.17.1-rc1 · apache/logging-log4j2,''
  \url{https://github.com/apache/logging-log4j2/releases/tag/log4j-2.17.1-rc1},
  (Accessed on 10/30/2024).

\bibitem{dependabot}
``Keeping your dependencies updated automatically with dependabot version
  updates - github docs,''
  \url{https://docs.github.com/en/code-security/dependabot/dependabot-version-updates},
  (Accessed on 10/30/2024).

\bibitem{Gertner_Zaromb_Schneider_Roberts_Matthews_2016}
\BIBentryALTinterwordspacing
A.~Gertner, F.~Zaromb, R.~Schneider, R.~Roberts, and G.~Matthews, ``The
  assessment of biases in cognition,'' Jun 2016. [Online]. Available:
  \url{https://www.mitre.org/news-insights/publication/assessment-biases-cognition}
\BIBentrySTDinterwordspacing

\bibitem{Khanmohammadi2023HalfDayVA}
\BIBentryALTinterwordspacing
K.~Khanmohammadi and R.~Khoury, ``Half-day vulnerabilities: A study of the
  first days of cve entries,'' \emph{ArXiv}, vol. abs/2303.07990, 2023.
  [Online]. Available: \url{https://api.semanticscholar.org/CorpusID:257505485}
\BIBentrySTDinterwordspacing

\bibitem{frei-similarstudies-1}
S.~Frei, M.~May, U.~Fiedler, and B.~Plattner, ``Large-scale vulnerability
  analysis,'' in \emph{Proceedings of the 2006 SIGCOMM Workshop on Large-Scale
  Attack Defense}, ser. LSAD '06.\hskip 1em plus 0.5em minus 0.4em\relax New
  York, NY, USA: Association for Computing Machinery, 2006, p. 131–138.

\bibitem{frei-similarstudies-2}
S.~Frei, D.~Schatzmann, B.~Plattner, and B.~Trammell, ``Modeling the security
  ecosystem - the dynamics of (in)security,'' in \emph{Economics of Information
  Security and Privacy}, T.~Moore, D.~Pym, and C.~Ioannidis, Eds.\hskip 1em
  plus 0.5em minus 0.4em\relax Boston, MA: Springer US, 2010, pp. 79--106.

\bibitem{effort_estimate}
T.~Sommestad, H.~Holm, and M.~Ekstedt, ``Effort estimates for vulnerability
  discovery projects,'' in \emph{2012 45th Hawaii International Conference on
  System Sciences}, 2012, pp. 5564--5573.

\bibitem{mitre_bias}
S.~Christey and B.~Martin, ``Buying into the bias: Why vulnerability statistics
  suck,'' \emph{BlackHat, Las Vegas, USA, Tech. Rep}, vol.~1, 2013.

\bibitem{right_source_vulnerability}
F.~Massacci and V.~H. Nguyen, ``Which is the right source for vulnerability
  studies? an empirical analysis on mozilla firefox,'' in \emph{Proceedings of
  the 6th International Workshop on Security Measurements and Metrics}, ser.
  MetriSec '10.\hskip 1em plus 0.5em minus 0.4em\relax New York, NY, USA:
  Association for Computing Machinery, 2010.

\bibitem{vulnerability_prediction}
\BIBentryALTinterwordspacing
V.~H. Nguyen and L.~M.~S. Tran, ``Predicting vulnerable software components
  with dependency graphs,'' in \emph{Proceedings of the 6th International
  Workshop on Security Measurements and Metrics}, ser. MetriSec '10.\hskip 1em
  plus 0.5em minus 0.4em\relax New York, NY, USA: Association for Computing
  Machinery, 2010. [Online]. Available:
  \url{https://doi.org/10.1145/1853919.1853923}
\BIBentrySTDinterwordspacing

\bibitem{ozment_vulnerability}
J.~A. Ozment, ``Vulnerability discovery \& software security,'' 2007.

\bibitem{malaiya_prediction}
O.~Alhazmi and Y.~Malaiya, ``Prediction capabilities of vulnerability discovery
  models,'' in \emph{RAMS '06. Annual Reliability and Maintainability
  Symposium, 2006.}, 2006, pp. 86--91.

\end{thebibliography}

\end{document}